\input phyzzx 
\font\boldsym=cmmib12\skewchar\boldsym='177
\textfont9=\boldsym \scriptfont9=\boldsym \scriptscriptfont9=\boldsym

\mathchardef\xib="0918
\def\sech{{\rm sech}}


\catcode`\@=11 
\def\eqnameb#1{\rel@x {\pr@tect
  \ifnum\equanumber<0 \xdef#1{{\rm(\number-\equanumber)}}%
     \gl@bal\advance\equanumber by -1
  \else \gl@bal\advance\equanumber by 1
   \xdef#1{{\rm(\ifcn@@ \chapterlabel.\fi \number\equanumber}}\fi
  }}

\baselineskip=18pt 
\parskip=0pt
\hoffset=0.2truein
\hsize=6truein
\voffset=0.1truein
\def\TITLEPAGE{\frontpagetrue}
\def\ITP#1{\hbox to\hsize{\tenpoint \baselineskip=12pt
        \hfil\vtop{
        \hbox{\strut ITP-SB-97-#1}
}}}
\def\DAMTPH{\hbox to\hsize{\tenpoint \baselineskip=12pt
        \hfil\vtop{
        \hbox{\strut DAMTP-1998-162}
}}}
\def\PRINCETON{
\centerline{\tenpoint Department of Astrophysical Sciences, %
Princeton University, 
Princeton, New Jersey 08544, USA\footnote\ddagger{E-mail: %
dns@astro.princeton.edu}}
}

\def\SB{
\centerline{\tenpoint Institute for Theoretical Physics, %
State University of New York, 
Stony Brook, New York 11794, USA}}

\def\DAMTP{
\centerline{\tenpoint DAMTP, University of Cambridge, Silver Street, 
Cambridge CB3 9EW, UK\footnote\dagger{Current address. %
E-Mail: M.A.Bucher@damtp.cam.ac.uk}}
}

\def\TITLE#1{\vskip .0in \centerline{\fourteenpoint #1}}
\def\AUTHOR#1{\vskip .1in \centerline{#1}}

\def\ABSTRACT#1{\vskip .1in \vfil \centerline{\twelvepoint
\bf Abstract}
\vskip 10pt
   #1 \vfil}
\def\ENDTITLEPAGE{\vfil\eject\pageno=1}
\hfuzz=5pt
\tolerance=10000
\baselineskip=0.33in
\TITLEPAGE
\DAMTPH
\ITP{22}

\vskip 15pt

\TITLE{\fourteenbf Is the Dark Matter a Solid?}

\vskip 5pt 
{\baselineskip=14pt
\AUTHOR{Martin Bucher}
\DAMTP
\SB
}

{\baselineskip=14pt
\AUTHOR{David Spergel}
\PRINCETON
}

\ABSTRACT{
\baselineskip 14pt 
A smooth unclustered dark matter component with
negative presure could reconcile a flat universe
with the many observations that find a density
in ordinary, clustered matter well below the
critical density and also explain the recent
high redshift supernova data suggesting that the 
expansion of the universe is now accelerating. 
For a perfect fluid negative 
presure leads to instabilities that are
most severe on the shortest scales. However, if 
instead the dark matter is a solid, with an
elastic resistance to pure shear deformations, 
an equation of state with negative presure 
can avoid these short wavelength instabilities.
Such a solid may arise as the result of different
kinds of microphysics. Two possible candidates for a 
solid dark matter component are a frustrated network of 
non-Abelian cosmic strings or a frustrated network
of domain walls. If these networks
settle down to an equilibrium 
configuration that gets carried along and 
stretched by the Hubble
flow, equations of state result with $w=-1/3$ and 
$w=-2/3,$ respectively. 
One expects the sound speeds for the solid dark matter
component to comprise an appreciable 
fraction of the speed of light. Therefore,
the solid dark matter does not cluster,
except on the very largest scales, accessible
only through observing the large-angle CMB
anisotropy.  In this paper we develop
a generally-covariant, continuum description for the dynamics
of a solid dark matter component.
We derive the evolution equations for
the cosmological perturbations in a flat universe
with CDM+(solid) and compute the resulting 
large-angle CMB anisotropy. 
The formalism presented here applies to any
generalized dark matter with negative pressure and a non-dissipative
resistance to shear. 
}

\rightline{[November 1998---Revised May 1999]}

\ENDTITLEPAGE

\REF\bfan{N.~Bahcall and X.~Fan, ``A Lightweight Universe?"  
Proc. Nat. Acad. Sci. {\bf 95,} 5956 (1998)(astro-ph/9804082).}

\REF\book{E.P.S.~Shellard and A.~Vilenkin, {\it Cosmic Strings and Other
Topological Defects,} (Cambridge:~Cambridge University Press, 1994);
A.~Vilenkin, ``Cosmic Strings and Domain Walls," 
Phys.~Rept.~{\bf 121,} 263 (1985).}

\REF\reviews{N.D.~Mermin, ``The Topological Theory of Defects in Ordered
Media,'' Rev.~Mod.~Phys.~{\bf 51,} 591 (1979);
J.~Preskill, ``Vortices and Monopoles," in P.~Ramond and R.~Stora, Eds., 
{\it Architecture of the Fundamental Interactions at Short Distances,} 
(Amsterdam:~North Holland, 1987).}

\REF\eone{A.~Einstein, ``Kosmologische Betrachtungen zur allgemeinen
Relativit\"atstheorie," Preuss.~Akad.~Wiss.~Berlin, Stizber.~142 (1917).}

\REF\etwo{Quoted by G.~Gamow in {\it My World Line,} (New York:
Viking, 1970) p. 44.}

\REF\ccnew{L.~Kofman and A.A.~Starobinskii, ``Effect of
the cosmological constant on large-scale anisotropies in the microwave background", Soviet~Astron.~Letters, {\bf 11}, 271 (1984)}

\REF\weinberg{ S. ~Weinberg, ``Theories of the Cosmological
Constant,'' in N.~Turok, Ed.,
{\it Critical Dialogues in Cosmology,} (Singapore, World Scientific,
1997); S.~Weinberg, ``The Cosmological Constant Problem,"
Rev.~Mod.~Phys.~{\bf 61,} 1 (1989).}

\REF\kochanek{E.~Falco, C.S.~Kochanek and J.A.~Munoz, ``Limits on 
Cosmological Models from Radio-Selected Gravitational Lenses,"
Ap.J. 494, 47 (1998); C.S.~Kochanek, ``Is There a Cosmological
Constant?", Ap.J. 466, 638 (1996).
}

\REF\garnavich{Garnavich, P.M. et al.~1998a, "Constraints on Cosmological 
Models from Hubble Space Telescope Observations of High-z Supernovae,"
Ap.J. (Lett.) 493, L53; Riess et al. 1998, ``Observational Evidence 
from Supernovae for an Accelerating Universe and a Cosmological Constant,"
A.J. 116, 1009 (astro-ph/9805201);
Garnavich, P.M. et al.~1998b, ``Supernova Limits on the Cosmic
Equation of State," Ap.J. {\bf 509,} 74 (astro-ph/9806396)
}

\REF\perlmutter{Perlmutter, S.~et al.~(1998),
``Measurements of Omega and Lambda from 42 High-Redshift Supernovae,"
(astro-ph/9812133) (to appear in Ap.J.);
Perlmutter, S.~et al.~(1998), ``Discovery of a Supernova Explosion
at Half the Age of the Universe and its Cosmological Implications,"
Nature, 391, 51 (astro-ph/9712212)}

\REF\vilenkin{
T.W.B.~Kibble, ``Topology of Cosmic Domains and Strings,"
J.~Phys.~{\bf A9,} 1387 (1976); A.~Vilenkin, ``Cosmological Density
Fluctuations Produced by Vacuum Strings," Phys.~Rev.~Lett.~{\bf 46,}
1169 (1981) [Erratum: Ibid.~{\bf 46,} 1496].
}

\REF\scal{
T.W.B.~Kibble, ``Topology of Cosmic Strings and Domains,"
J.~Phys.~{\bf A9,} 1387 (1976);
T.W.B.~Kibble, ``Some Implications of a Cosmological Phase 
Transition," Phys.~Rep.~{\bf 67,} 183 (1980);
A.~Vilenkin, ``Cosmic Strings," Phys.~Rev.~{\bf D24,} 2082 (1981).}

\REF\numscal{
A.~Albrecht and N.~Turok, ``Evolution of Cosmic Strings,"
Phys.~Rev.~Lett.~{\bf 54,} 1868 (1985);
A.~Albrecht and N.~Turok, ``Evolution of Cosmic String Networks,"
Phys.~Rev.~{\bf D40,} 973 (1989);
D.~Bennett and F.~Bouchet, ``Evidence for a Scaling Solution in
Cosmic String Evolution," Phys.~Rev.~Lett.~{\bf 60,} 257 (1988);
D.~Bennett and F.~Bouchet, ``High Resolution Simulations of 
Cosmic String Evolution: Network Evolution," Phys.~Rev.~{\bf D41,}
2408 (1990);
B.~Allen and E.P.S.~Shellard, ``Cosmic String Evolution---A 
Numerical Simulation," Phys.~Rev.~Lett.~{\bf 64,} 119 (1990).}

\REF\inter{
E.P.S.~Shellard, ``Cosmic String Interactions," Nucl.~Phys.~{\bf B283,}
624 (1987); R.A.~Matzner, ``Interactions of $U(1)$ Cosmic Strings:
Numerical Intercommutation," Computers in Physics, Sept./Oct.~51, (1988);
K.~Moriarty, E.~Meyers and C.~Rebbi, ``Dynamical Interactions of 
Cosmic Strings and Flux Vortices in Superconductors," Phys.~Lett.~{\bf 207B,}
411 (1988).}

\REF\poen{V.~Poenaru and G.~Toulouse, ``The Crossing of Defects in Ordered
Media and the Topology of 3-Manifolds," J.~Phys.~(Paris) {\bf 38,}
887 (1977); F.~Bais, ``Flux Metamorphosis," 
Nucl.~Phys.~{\bf B170,} 32 (1980).}

\REF\bucher{M.~Bucher, ``The Aharonov-Bohm Effect and Exotic Statistics
for Non-Abelian Vortices," Nucl.~Phys.~{\bf B350,} 163 (1991).} 

\REF\pen{D.~Spergel and U.~Pen, ``Cosmology in a String-Dominated Universe,"
Ap.~J.~(Letters) {\bf 491,} L67 (1997) (astro-ph 96-11198).}

\REF\mcgraw{
P.~McGraw, ``Evolution of a Nonabelian Cosmic String Network,"
Phys.~Rev.~{\bf D57,} 3317 (1998) (astro-ph/9706182);
P.~McGraw, ``Dynamical Simulation of Nonabelian Cosmic Strings,"
(hep-th 96-03153) (March 1996).}

\REF\vtwo{A.~Vilenkin, ``String Dominated Universe," 
Phys.~Rev.~Lett.~{\bf 53,} 1016 (1984).}

\REF\vv{T. Vachaspati and A.~Vilenkin, ``Evolution of Cosmic Networks,"
Phys.~Rev.~{\bf D35,} 1131 (1987).}

\REF\landau{L.~Landau and E.~Lifshitz, {\it Theory of Elasticity,} Third 
Edition, (New York, Pergamon, 1986).}

\REF\ks{H.~Kodama and M.~Sasaki, ``Cosmological Perturbation Theory,"
Prog.~Theor.~Phys.~(Suppl.) {\bf 78,} 1 (1984).}

\REF\dwr{B.~Ryden, W.~Press and D.~Spergel, ``The Evolution of Networks of 
Domain Walls and Cosmic Strings," Ap.~J.~{\bf 357,} 293 (1990)}

\Ref\kubo{H. Kubotani, 
``The Domain Wall Network of Explicitly Broken $O(N)$ Model,"
Prog.~Theor.~Phys.~{\bf 87,} 387 (1992).}

\REF\quintr{B.~Ratra and P.J.E.~Peebles, ``Cosmological 
Consequences of a Slowly Rolling Scalar Field,"
Phys.~Rev.~{\bf D37,} 3406 (1988); 
J.~Frieman, C.~Hill, A.~Stebbins and I.~Waga,
``Cosmology with Ultralight Psuedo Nambu Goldstone Bosons,"
Phys. Rev. Lett. {\bf 75,} 2077 (1995); 
P.~Vaina and A.~Liddle, ``Perturbation Evolution in Cosmologies 
with a Decaying Cosmological Constant," Phys.~Rev.~{\bf D57,} 674 (1998)
(astro-ph/9708247); 
K.~Coble, S.~Dodelson and J.~Frieman , ``Dynamical Lambda Models of Structure
Formation," Phys.~Rev.~{\bf D55,} 1851 (1997) (astro-ph/9608122);
R.~Caldwell, R.~Dave and P.~Steinhardt, ``Cosmological Imprint of an 
Energy Component with General Equation of State,"
Phys.~Rev.~Lett.~{\bf 80,} 1582 (1998);
M. Turner and M. White, ``CDM Models with a Smooth Component,"
Phys. Rev. D56, 4439 (1997) (astro-ph/9701138);
T. Chiba, N. Sugiyama and T. Nakamura, ``Cosmology with X Matter,"
MNRAS {\bf 289,} L5 (1997) (astro-ph/9704199);
T. Chiba, N. Sugiyama and T. Nakamura,
"Observational Tests of X Matter Models," 
MNRAS {\bf 301,} 72 (1998)
(astro-ph/9806332).}
 
\REF\cpma{C.P.~Ma and E.~Bertschinger, ``Cosmological
Perturbation Theory in the Synchronous and Conformal Newtonian Gauges," 
Ap.~J.~{\bf 471,} 13 (1996) (astro-ph/9605198).}

\REF\feld{V.~Mukhanov, H.~Feldman and R.~Brandenberger,
``Theory of Cosmological Perturbations," Phys.~Repts.~{\bf 215,}
203 (1992).}

\REF\hu{W.~Hu, ``Structure Formation with Generalized Dark Matter,"
Ap.J. {\bf 506,} 485 (1998) (astro-ph/9801234).}

\REF\avthree{A. Vilenkin, private communication.}

\REF\carter{B. Carter and H. Quintana, Proc. Roy. Soc. Lond. 
{\bf A331,} 57 (1972);
B.~Carter, ``Elastic Perturbation Theory in General Relativity
and a Variation Principle for a Rotating Solid Star,"
Comm. Math. Phys. {\bf 30,} 261 (1973); B. Carter, ``Interaction
of Gravitational Waves with an Elastic Medium," in
T. Piran and N. Deruelle, Eds., {\it Rayonnement Graviationel,}
(Amsterdam: North Holland, 1983).}

\REF\gretother{B. DeWitt, ``The Quantization of Geometry,"
in L. Witten, {\it Gravitation: An Introduction to Current
Research,} (New York, Wiley) (1962); J. Souriau, 
{\it Geometrie et Relativit\'e} (Paris, Hermann) (1964);
C. Cattaneo, "Elasticit\'e Relativiste," Symp. Math. {\bf 12,}
337 (1973); G.A. Maugin, ``On the Covariant Equations of the
Relativistic Electrodynamics of Continua III: Elastic 
Solids," J. Math, Phys. {\bf 19,} 1212 (1978);
J. Kijowski and G. Magli, ``Unconstrained Hamiltonian
Formulation of General Relativity with Thermo-Elastic Sources,"
Class. Quant. Grav. {\bf 15,} 3891 (1998).}

\REF\eichler{D.~Eichler, ``Condensed Dark Matter?,"
Ap.~J.~468, 75 (1996).}

\baselineskip=16pt

\def\y{{\bf y}}
\def\H{{\cal H}}

\def\hc{{\tt h}}
\def\D{{\cal D}}

\chapter{\bf Introduction}

Most determinations of the cosmological density parameter
$\Omega _0=(\rho /\rho _{crit}),$ where $\rho _{crit}= (3/8\pi G)~H_0^2,$
now indicate that $\Omega _m\approx 0.2\pm 0.1,$ 
a value well below the $\Omega _m=1$ value 
suggested by flat cosmological models. (For a nice review of 
the current observations see ref.~\bfan .) Most of these techniques
for determining $\Omega _m,$ however, are sensitive only to matter
that is clustered gravitationally and do not rule out a smooth,
unclustered component that could make up the difference between
the observed value of $\Omega _m$ and unity.

The earliest proposal for a smooth component is the cosmological
constant $\Lambda ,$ first introduced by Einstein,\refmark{\eone }
later denounced by him,\refmark{\etwo }
and more recently resurrected to reconcile the observations with
a flat universe.\refmark{\ccnew }
The cosmological constant is somewhat of an
embarrassment for theoretical physics because dimensional arguments
would suggest $\Lambda \approx {M_{pl}}^4,$ a value more than 
a hundred orders of magnitude too big!\refmark{\weinberg }
Perhaps some
not yet discovered symmetry makes $\Lambda $ vanish exactly,
but at this point we lack even the vaguest idea of what kind of
symmetry could do the job. Supersymmetry somewhat mitigates 
the difficulty, making ${M_{SSB}}^4$ rather than ${M_{pl}}^4$ the
naive guess for $\Lambda ,$ but even with supersymmetry 
if $\Lambda $ does not vanish, a formidable fine tuning problem
persists. A large $\Omega _{\Lambda }$ overpredicts the number 
of gravitationally lensed quasars.\refmark{\kochanek }
As an alternative to $\Lambda ,$ it has been proposed that there could exist 
a very light, extremely weakly coupled scalar field that could act as a 
temporary cosmological constant, even though the true value of 
the cosmological constant vanishes exactly.\refmark{\quintr }
But this requires a particle of 
implausibly small mass, somewhere in the neighborhood of
${10}^{-33}~{\rm eV}.$ 

In this paper we discuss another possibility: a solid dark 
matter component with significant negative pressure. 
Here {\it significant} means that the negative pressure,
or equivalently tension, of the solid matter component is 
comparable in magnitude to its energy density. 
An equation of state with large negative pressures 
can lead to sound speeds comparable to the speed of
light, so that the Jeans length for this component
is enormous, comparable to the size of our present
horizon. Consequently, the solid dark matter component
does not cluster except on extremely large scales.
Because of this the low measurements of $\Omega $
can be reconciled with a spatially flat universe.
The clustering of the solid dark matter component
on very large scales is accessible to observation 
only through its effect on the large-angle CMB anisotropy. 

A solid dark matter component can also help explain the 
the recent observations of distant Type Ia supernovae
that suggest that the universe is now 
accelerating.\refmark{\garnavich , \perlmutter } 
For the expansion of the universe to accelerate
some exotic form of matter with $w=(p/\rho )<-1/3$
is required. A perfect fluid with $w<0$ is not
possible because its sound speed would be 
imaginary, indicating instabilities whose
growth rate diverges as the wavelength 
approaches zero. Such a fluid would clearly be 
unphysical. For a solid, however, real sound
speeds are possible because a shear
modulus of sufficient magnitude removes these
instabilities. 

In this paper we explore the dynamics of a solid
dark matter component by developing a continuum
description for such a component within the 
framework of general relativity and incorporating
the solid dark matter component into the linearized theory
for the evolution of cosmological perturbations.
In particular we explore the consequences of 
such a component for the large angle CMB 
anisotropy.

A solid dark matter component could arise 
from a variety of different microphysics.
Two known ways such a component could
arise are from networks of frustrated
cosmic strings\refmark{\vtwo ,\mcgraw , \pen} 
or domain walls.\refmark{\kubo , \dwr} 
While the simplest Abelian cosmic strings
obey a scaling solution so that the number
of strings per horizon volume remains constant,
for non-Abelian cosmic strings topological
obstructions prevent the intercommuting 
necessary for the breakup long strings that
leads to scaling behavior. The nonunit elements
of the fundamental group $\pi _1(G/H)$ classify
the possible types of cosmic strings.
When two strings whose windings or magnetic
fluxes are described by non-commuting elements
of $\pi _1(G/H)$ try to cross, the strings 
cannot pass through
each other without forming a third string between
them. This has the effect of preventing crossings
because the tension of the intermediate strings tries
to pull the two strings back toward their
previous uncrossed positions. It is possible that these
effects lead to a scaling solution albeit one
with many more strings per horizon volume, but the
simulations by Pen and Spergel suggest that the strings
settle down to a stable configuration
which subsequently gets carried along with the Hubble
flow. In a forthcoming paper, we show that
stable string configurations do exist which strengthens the case for
a string dominated universe. Similar simulations of domain walls 
by Kubotani suggest domain walls in models with many
types of domain walls exhibit similar behavior. 
A cellular foam type structure in equilibrium 
forms with several wall meeting at linelike 
junctions. 
A string-dominated universe gives $w=-1/3,$
which in the absence of other dark matter
would give a universe that is neither
accelerating or decelerating. A symmetry
breaking scale of a few TeV and a 
string separation today of a few A.U.
would give $\Omega _{string}$ today
in the interesting range between zero
and one. (This estimate is subject to
considerable uncertainty because a
range of string density at formation
is possible and the length of the 
transient before string dominated 
behavior takes over is uncertain and
model dependent).
For domain walls carried along with the
Hubble flow, $w=-2/3$ and a symmetry
breaking scale of a few MeV and 
a mean domain wall separation of some tens
of parsecs are suggested (subject to
the same uncertainties).
The fact that new physics occurs at 
larger energy scales than for
{\it quintessence} or $\Lambda $
is a positive feature of these scenarios.
It should be stressed that the formalism 
in the paper applies equally well to a
solid component with the same continuum 
description but of completely different
origin.

A slightly different type of solid dark
matter has been proposed by Eichler to
explain certain aspects of 
large-scale structure.\refmark{\eichler }
In this scenario a solid condenses
and subsequently fractures when 
stretched beyond its breaking point by 
expansion of the universe. 
The dark matter contemplated in this paper
does not fracture. It can experience unlimited
stretching without becoming in any sense 
weakened. For a {\it solid} composed of frustrated 
topological defects it is easy to see why ruptures or
fractures do not occur. The constituent
defects lack a preferred size. Upon
stretching or shrinking, the
transverse structure of the defects
remains unchanged. This is quite unlike
an ordinary  solid composed of atoms, for 
which quantum mechanics establishes a 
preferred length for the chemical bonds. 

The organization of this paper is as follows. 
In Sect.~2 of this paper we develop a generally 
covariant description of the dynamics of a 
continuous medium (such as the string network)
in curved space. For the spacetime with the metric
$G_{\mu \nu }=a^2(\eta )~[\eta _{\mu \nu }+\hc _{\mu \nu }]$
where $\hc _{\mu \nu }$ is regarded as small, we expand the
action to quadratic order and compute the equations
of motion and the resulting stress-energy for the solid dark
matter component.
In Sect.~3 we combine the results of Sect.~2 with the linearized
theory of cosmological perturbations using Newtonian
gauge and derive the equations of motion for a 
spatially flat universe with cold dark matter
(CDM) and a solid dark matter component.
In Sect.~4 compute the large-angle CMB anisotropy
for models with SDM. Finally, in Sect.~5 we present
some concluding remarks. In this paper we use the sign convention
$\eta _{\mu \nu }={\rm diag}[-1, +1, +1, +1].$

\chapter{\bf Continuum Description}

This section presents an action that describes the 
dynamics of a dissipationless
elastic medium in curved space. Although developed to
describe the response of a non-Abelian string network to 
metric cosmological perturbations, this formalism applies
to other forms of solid dark matter and to
a wide variety of situations involving continuous media
in curved space. The problem of describing the dynamics
of a solid within the framework of general relativity
has been previously considered by Carter and Quintana
in the study of the crusts of neutron stars\refmark{\carter }
and others.\refmark{\gretother } 
In this section we 
present a self contained treatment particularly suited to
the consideration of linearized perturbations in an 
expanding universe. 

A continuous medium is a kind of three-dimensional membrane but quite
different from the now much discussed `fundamental' branes.
A continuous medium has internal structure. As the medium
moves, each constituent particle traces out its own 
worldline. For continuous media the allowed reparameterizations 
are limited to reparameterizations that preserve these worldlines. 
Geometrically a continuous medium may be regarded 
as a congruence of worldlines. We use a three dimensional
coordinate $\y $ to parameterize these worldlines. The 
coordinate $t$ is an arbitrary time coordinate
parameterizing the direction along these worldlines.

A metric $h_{ab}$ is defined on $\y $-space. The volume
induced by this metric indicates
the density of worldlines and the additional
internal metric structure provides a reference with respect to which
to characterize pure shear (i.e., volume preserving) deformations.
In the classical exposition of elasticity theory (e.g. as in 
the book by Landau and Lifshitz\refmark{\landau })
$h_{ab}=\delta _{ab}=(constant).$ However, when a solid is formed
in a warped (i.e., curved) spacetime, there generally does not
exist any coordinatization of the worldlines so that 
$h_{ab}=\delta _{ab}.$ 
For the moment let us imagine ourselves
in the instantaneous rest frame of a volume element of the medium,
choosing $t$ so that $\partial /\partial t$ is orthogonal
to $\partial /\partial y^a$ ($a=1, 2, 3$). The background 
spacetime metric $G_{\mu \nu }$ induces the following metric on 
$\y $-space:
$$ g_{ab}=G_{\mu \nu }{\partial X^\mu \over \partial y^a}
{\partial X^\nu \over \partial y^b}.\eqn\maa $$
For an arbitrary time parameterization, where $\partial /\partial t$
is not necessarily orthogonal to $\partial /\partial y^a,$
the induced metric may rewritten as 
$$ g_{ab}=G_{\mu \nu }^{(s)}{\partial X^\mu \over \partial y^a}
{\partial X^\nu \over \partial y^b}\eqn\maab$$
where $G_{\mu \nu }^{(s)}=U_\mu U_\nu +G_{\mu \nu }$ and 
$U^\mu ={\partial X^\mu \over \partial t}\big/
\sqrt{G_{\xi \eta } {\partial X^\xi \over \partial t}
{\partial X^\eta \over \partial t} }.$
$G^{(s)}_{\mu \nu }$ projects out displacements along 
$\partial /\partial t.$ 

The local deformation state of the medium is determined by
comparing $g_{ab}$ to $h_{ab}$---by the three 
principal values $\lambda _1,  \lambda _2, \lambda _3$ 
of $g_{ab}$ with respect to $h_{ab}.$ (We assume that
the medium is isotropic, for otherwise more structure
than $h_{ab}$ alone is required to characterize the 
deformation state of the medium.)
In ordinary elasticity theory, $h_{ab}=\delta _{ab},$
$g_{ab}$ is the strain tensor, and $\lambda _1,$
$\lambda _2,$ $\lambda _3$ are its principal values.   
The scalar invariants 
$g_{(1)}=h^{ab}g_{ab},$ $g_{(2)}=h^{ab}g_{bc}h^{cd}g_{da},$
$g_{(3)}=h^{ab}g_{bc}h^{cd}g_{de}h^{ef}g_{fa}$ 
suffice to characterize completely the principal values,
and the local density in the local instantaneous rest frame
with respect to the $h$ volume element may 
be expressed as 
$ \rho _{(h)}= \rho _{(h)}(g_{(1)},g_{(2)},g_{(3)}).$
In terms of the principal values
$g_{(1)}={\lambda _1}+{\lambda _2}+{\lambda _3},$
$g_{(2)}={\lambda _1}^2+{\lambda _2}^2+{\lambda _3}^2,$
and 
$g_{(3)}={\lambda _1}^3+{\lambda _2}^3+{\lambda _3}^3.$

It follows that the action is 
$$ S=-\int dt \int d^3y~\sqrt{h}~
\rho \bigl( g_{(1)},  g_{(2)},  g_{(3)} \bigr) 
\sqrt{-G_{\mu \nu }
{\partial X^\mu \over \partial t}
{\partial X^\nu \over \partial t}}.\eqn\mact$$
One may view \mact~as a generalization of the free
particle action $S=-m\int d\tau .$ If $\rho (g_{(1)}, g_{(2)}, 
g_{(3)} )=
{\rm (constant)},$ the action \mact ~merely
describes a congruence of
non-interacting, freely-falling particles. However, 
in the general case the co-moving density with
respect to the internal coordinates varies as deformations
of the medium alter its internal energy. The potential
energy of the medium resides in the function 
$\rho (g_{(1)}, g_{(2)}, g_{(3)}).$ 

We now recast the action \mact ~into a more familiar form by
considering an elastic medium in almost flat space. We assume
a spacetime metric $G_{\mu \nu }=\eta _{\mu \nu }+\hc _{\mu \nu },$
an internal metric 
$h_{ab}=\delta _{ab}+b_{ab},$ and $X^i=y^i+\xi ^i(\y , t)$
where $\hc _{\mu \nu },$ $b_{ab},$ and $\xi ^i(\y , t)$
are regarded as small. We also set $X^0=t.$

We  expand
$$ \rho =\rho _s+\tau _s\left( {\delta V\over V}\right) 
+K_s~\left( {\delta V\over V}\right) ^2 +
\mu _s~S_{(5)ab}~S_{(5)}^{ab}.\eqn\maad$$
Here $\tau _s$ is the tension and $K_s$ and $\mu _s$ are the
compressional and shear moduli, respectively, and 
the tensor $S_{(5)ab}$ is the pure shear component of
the strain tensor. 
{}From the relation (valid to quadratic order)
$$
\left( 1+ {\delta V\over V}\right) 
={\sqrt{ \vert g_{ab}\vert }\over 
\sqrt{ \vert h_{ab}\vert }}
={\sqrt{ \vert \delta _{ab}+s_{ab}\vert }\over 
\sqrt{ \vert \delta _{ab}+b_{ab}\vert }}
={
1+{s\over 2}+{s^2\over 8}-{s_{ij}s^{ij}\over 4}
\over 
1+{b\over 2}+{b^2\over 8}-{b_{ij}b^{ij}\over 4}
},\eqn\maae$$
it follows that
$${\delta V\over V}={s\over 2}
+{s^2\over 8}-{s_{ij}s^{ij}\over 4}
-{b\over 2} +{b_{ij}b^{ij}\over 4}+{b^2\over 8}-{b s\over 4}.
\eqn\maaf$$
Here to quadratic order
$$\eqalign{
s_{ab}&=
\Bigl\{  {\delta ^i}_a+{\xi ^i}_{,a}(\y , t )\Bigr\} 
\Bigl\{ \delta _{ij}+\hc _{ij}(X) 
+(\dot \xi _i+\hc _{0i})(\dot \xi _j +\hc _{0j})\Bigr\}
\Bigl\{ {\delta ^j}_b+{\xi ^j}_{,b}(\y , t )\Bigr\} -\delta _{ab}\cr
&= \xi _{a, b}+\xi _{b, a} +\hc _{ab} +\xi ^k\nabla _k ~\hc _{ab}
+\xi _{i, a}\xi _{i, b}+\hc _{ai}\xi _{i, b}
+\hc _{bi}\xi _{i, a}+\dot \xi _a~\dot \xi _b\cr 
&+\hc _{0a}\dot \xi _b+\hc _{0b}\dot \xi _a+\hc _{0a}\hc _{0b},}\eqn\maag $$
and to quadratic order
$$\eqalign{
{\delta V\over V}&=
\xi _{a, a}+{1\over 2}\hc _{aa}
+{1\over 2}\xi ^c\nabla _c~\hc _{aa}
-{1\over 2}\xi _{a,b}\xi _{b,a}
+{1\over 2}\xi _{a,a}\xi _{b,b}
-{1\over 2}\xi _{a,a}b_{bb}-{1\over 4}h_{aa}b_{bb}\cr 
&+{1\over 8}\hc _{aa}\hc _{bb} 
-{1\over 4}\hc _{ab}\hc _{ab} +{1\over 2}\xi _{a,a}\hc _{bb}
-{1\over 2}b_{aa}+{1\over 4}b_{ab}b_{ab} +{1\over 8}b_{aa}b_{bb}
+{1\over 2}\dot \xi _a~\dot \xi _a \cr 
&+\hc _{0a}\dot \xi _a+{1\over 2}\hc _{0a}\hc _{0b}.\cr 
}\eqn\maah$$
To linear order (which is sufficient here), the pure shear part
of the strain is
$$S_{ab}^{(5)}=\xi _{a,b}+\xi _{b,a}+\hc _{ab}-b_{ab}
-{1\over 3}\delta _{ab}~\bigl( ~2~\xi _{c,c}+\hc _{cc}-b_{cc} \bigr) .
\eqn\maal$$
Combining and expanding to quadratic order, we obtain 
$$\eqalign{S=&\int dt \int d^3y~
\left[ 1+{b\over 2}+{b^2\over 8} -{b_{ab}b^{ab} \over 4}\right] \cr 
&\times \Biggl[ \rho _s
+\tau _s\Bigl\{ 
\xi _{a, a}+{1\over 2}\hc _{aa}
+{1\over 2}\xi ^c\nabla _c~\hc _{aa} 
-{1\over 2}\xi _{a,b}\xi _{b,a}
+{1\over 2}\xi _{a,a}\xi _{b,b} 
-{1\over 2}\xi _{a,a}b_{bb}\cr 
&-{1\over 4}h_{aa}b_{bb}
+{1\over 8}\hc _{aa}\hc _{bb}
-{1\over 4}\hc _{ab}\hc _{ab} +{1\over 2}\xi _{a,a}\hc _{bb}
-{1\over 2}b_{aa}+{1\over 4}b_{ab}b_{ab}+{1\over 8}b_{aa}b_{bb}\cr 
&+ {1\over 2}\dot \xi _a~\dot \xi _a
+\hc _{0a}\dot \xi _a+{1\over 2}\hc _{0a}\hc _{0b}
\Bigr\} \cr 
&\hskip .5in +K_s\Bigl\{ {\xi ^i}_{,i}+
{1\over 2}{\hc ^i}_i-{1\over 2}{b^i}_i \Bigr\} ^2\cr 
&\hskip .5in +\mu _s\Bigl\{  \xi _{a,b}+\xi _{b,a}+\hc _{ab}-b_{ab}
-{1\over 3}\delta _{ab}~
\bigl( 2~\xi _{c,c}+\hc _{cc}-b_{cc}\bigr) \Bigr\} ^2 \Biggr] \cr 
&\times
\Biggl[ {\hc _{00}\over 2}
+{\hc _{00}^2\over 8}
+{1\over 2}\xi ^i\nabla _i~\hc _{00}+\hc _{0i}\dot \xi ^i
+{1\over 2}\dot \xi ^{i~2}-1\Biggr] ,\cr }
\eqn\maam$$
and expanding out to quadratic order and omitting 
terms that do not contribute to the equations of motion
or to the stress-energy $T^{\mu \nu },$ we obtain 
$$\eqalign{
S&=\int dt \int d^3y~~~\Biggl[ ~~
\left( {\rho _s-\tau _s\over 2}\right) \dot \xi ^2 
+(\rho _s-\tau _s)~\hc _{0i}\dot \xi ^i \cr
& -\rho _s~\Bigl\{ 
-{\hc _{00}\over 2}
-{b_{aa}\hc _{00}\over 4}
-{\hc _{00}^2\over 8}-{1\over 2}(\xi ^c\nabla _c) \hc _{00}\Bigr\} \cr 
&-\tau _s~\Bigl\{ 
\xi _{a,a}
+{1\over 2}\hc _{aa}
+{1\over 2}(\xi ^c\nabla _c) \hc _{aa}
-{1\over 2}\xi _{a,b}~\xi _{b,a}
+{1\over 2}\xi _{a,a}~\xi _{b,b}
-{1\over 2}\xi _{a,a}~\hc _{00}\cr 
&+{1\over 2}\xi _{a,a}~\hc _{bb}
+{1\over 4}\hc _{00}~b_{aa}
-{1\over 4}\hc _{00}~\hc _{aa}
+{1\over 8}\hc _{aa}~\hc _{bb} 
-{1\over 4}\hc _{ab}~\hc _{ab}
+{1\over 2}\hc _{0a}~\hc _{0a}
\Bigr\} \cr 
&-K_s~\Bigl\{ {\xi ^i}_{,i}+
{1\over 2}{\hc ^i}_i-{1\over 2}{b^i}_i \Bigr\} ^2\cr
&-\mu _s~\Bigl\{  \xi _{a,b}+\xi _{b,a}+\hc _{ab}-b_{ab}
-{1\over 3}\delta _{ab}~
\bigl( ~2~\xi _{c,c}+\hc _{cc}-b_{cc}\bigr) \Bigr\} ^2 \Biggr] .\cr
}\eqn\maama$$
It follows that the equation of motion for $\xib $ is
$$\eqalign{
&(\rho _s-\tau _s)\ddot \xi _i
+(\rho _s-\tau _s)\dot \hc _{0i}-
{1\over 2}(\rho _s -\tau _s)\nabla _i\hc _{00}
-K_s\Bigl[ 2\xi _{j,ji}+ \hc _{jj,i}- b_{jj,i}\Bigr] \cr
&~~~~~-\mu _s\Bigl[ 4\xi _{i,jj}+{4\over 3}\xi _{j,ij}
-4b_{ij,j}+{4\over 3}b_{jj,i}
+4\hc _{ij,j}-{4\over 3}\hc _{jj,i}\Bigr] =0.\cr }\eqn\maamb$$
With the coupling to gravity ignored, which is a reasonable 
approximation for short wavelengths, the sound speeds
$$ 
{c_S}^2={{16\over 3}\mu _s+2K_s\over (\rho _s-\tau _s)},\qquad 
{c_V}^2={4\mu _s\over (\rho _s-\tau _s)}
\eqn\grr$$
follow for the longitudinal (scalar) mode and the two 
transverse (vector) modes, respectively. 

We next compute the stress-energy, defined by the relation
$$\delta S={1\over 2}\int d^4x~\sqrt{-G}~T^{\mu \nu }~(\delta G_{\mu \nu }).
\eqn\maamc$$
We use the expansion 
$1/\sqrt{-\vert G_{\mu \nu }\vert }
=\bigl[ 1+\hc _{00}/2-\hc _{ii}/2+O(\hc ^2)\bigr] $
and the relation $A~w_{aa}^2+B(w_{ab}-{1\over 3}\delta _{ab}~w_{cc})^2
=(A-{1\over 3}B)~{w_{aa}}^2  +B~w_{ab}w_{ab}.$
It follows that (to linear order) 
$$\eqalign{
{T_0}^0&=-\rho _s -{1\over 2}(\rho _s-\tau _s)(b_{ii}-\hc _{ii}-2\xi _{i,i}),\cr
{T_i}^0&= (\rho _s-\tau _s)~(\dot \xi ^i+\hc _{0i}),\cr 
{T_i}^j&= -\tau _s~\delta _{ij}\cr 
&~~~-  2\left( K_s-{4\over 3}\mu _s\right)
\delta _{ij}~
\left( \xi _{a,a}+{1\over 2}\hc _{aa}-{1\over 2}b_{aa}\right) 
-4\mu _s\Bigl( \xi _{i,j}+\xi _{j,i}+\hc _{ij}-b_{ij}\Bigr) .\cr  
}\eqn\maamde$$

We now turn to the continuum description of the 
solid dark matter component with
the expansion of the universe included using
conformal time $\eta $ so that the background metric becomes
$$G_{\mu \nu }=a^2(\eta )\cdot 
\bigl[ \eta _{\mu \nu }+\hc _{\mu \nu }\bigr] .\eqn\groga$$
We modify the notation for the expansion of the 
co-moving density as follows
$$\rho =\rho _s~a^\lambda (\eta )\left\{ 1+
\tilde \tau _s\left( {\delta V\over V}\right) 
+\tilde K_s\left( {\delta V\over V}\right) ^2
+\tilde \mu _s~S_{(5)ab}~{S_{(5)}}^{ab} \right\} ,\eqn\maan$$
so that $\tilde \tau _s,$ $\tilde K_s,$ and $\tilde \mu _s$ are 
dimensionless. We take the exponent $\lambda $ to be constant,
although the generalization of this is straightforward. 
The physical density scales as $\rho _{phys}=\rho _{cm}/a^3(\eta)
=\rho _s~a^{\lambda -3}(\eta ).$ Given
$\lambda ,$ the dimensionless parameters 
$$ \tilde \tau _s ={\lambda \over 3},\qquad 
\tilde K_s ={\lambda (\lambda -3)\over 18}
\eqn\grra$$
are fixed, and $\tilde \mu _s$ is variable only within the range
$$ {\rm max}\Bigl[ ~0, {-3\over 8}\tilde K_s~\Bigr] 
\le \tilde \mu _s \le {(1-\tilde \tau _s)\over 4},
\eqn\grrb $$
obtained by requiring stability on short wavelengths and 
subluminal longitudinal and transverse sound speeds.
For frustrated non-Abelian strings, $\tilde \tau _s=1/3$ and 
$\tilde K_s=-1/9;$ for a frustrated network of domain walls,
$\tilde \tau _s=2/3$ and $\tilde K_s=-1/9.$  

For the expanding universe, 
the action \maam ~is modified as follows:
$$\eqalign{S=&\int d\eta  \int d^3y~\rho _s~a^{\lambda +1}(\eta )
\left[ 1+{b\over 2}+{b^2\over 8} -{b_{ab}b^{ab} \over 4}\right] \cr
&\times \Biggl[ 1
+\tilde \tau _s\Bigl\{
\xi _{a, a}+{1\over 2}\hc _{aa}
+{1\over 2}\xi ^c\nabla _c~\hc _{aa}
-{1\over 2}\xi _{a,b}\xi _{b,a}
+{1\over 2}\xi _{a,a}\xi _{b,b}
-{1\over 2}\xi _{a,a}b_{bb}\cr
&-{1\over 4}h_{aa}b_{bb}
+{1\over 8}\hc _{aa}\hc _{bb}
-{1\over 4}\hc _{ab}\hc _{ab} +{1\over 2}\xi _{a,a}\hc _{bb}
-{1\over 2}b_{aa}+{1\over 4}b_{ab}b_{ab}+{1\over 8}b_{aa}b_{bb}\cr
&+ {1\over 2}\dot \xi _a~\dot \xi _a
+\hc _{0a}\dot \xi _a+{1\over 2}\hc _{0a}\hc _{0b}
\Bigr\} \cr
&\hskip .5in +\tilde K_s\Bigl\{ {\xi ^i}_{,i}+
{1\over 2}{\hc ^i}_i-{1\over 2}{b^i}_i \Bigr\} ^2\cr
&\hskip .5in +\tilde \mu _s\Bigl\{  \xi _{a,b}+\xi _{b,a}+\hc _{ab}-b_{ab}
-{1\over 3}\delta _{ab}~
\bigl( 2~\xi _{c,c}+\hc _{cc}-b_{cc}\bigr) \Bigr\} ^2 \Biggr] \cr
&\times
\Biggl[ {\hc _{00}\over 2}
+{\hc _{00}^2\over 8}
+{1\over 2}\xi ^i\nabla _i~\hc _{00}+\hc _{0i}\dot \xi ^i
+{1\over 2}\dot \xi ^{i~2}-1\Biggr] .\cr }
\eqn\moao$$
The equations of motion  are modified to 
$$\eqalign{
&(1-\tilde \tau _s)\bigl[ \ddot \xi _i +
(\lambda +1){\dot a\over a}\dot \xi _i\bigr]
+(1-\tilde \tau _s)\bigl[ \dot \hc _{0i}+
(\lambda +1){\dot a\over a}\hc _{0i}\bigr]-
{1\over 2}(1-\tilde \tau _s)\nabla _i\hc _{00}\cr
& -\tilde K_s\Bigl[ 2\xi _{j,ji}+ \hc _{jj,i}- b_{jj,i}\Bigr] 
-\tilde \mu _s\Bigl[ 4\xi _{i,jj}+{4\over 3}\xi _{j,ij}
-4b_{ij,j}+{4\over 3}b_{jj,i}
+4\hc _{ij,j}-{4\over 3}\hc _{jj,i}\Bigr] =0.\cr }
\eqn\moabb$$
In the expanding universe,
the stress-energy (to linear order) is
$$\eqalign{
{T_0}^0&=a^{(\lambda -3)}\rho _s\Bigl[ -1 -{1\over 2}
(1-\tilde \tau _s)(b_{ii}-\hc _{ii}-2\xi _{i,i}) \Bigr] \cr
{T_i}^0&=a^{(\lambda -3)}\rho _s\Bigl[ (1-\tilde\tau _s)~
(\dot \xi ^i+\hc _{0i}) \Bigr] ,\cr
{T_i}^j&=a^{(\lambda -3)}\rho _s\Bigl[ -\tilde\tau _s~\delta _{ij}
-  2\left( \tilde K_s-{4\over 3}\tilde\mu _s\right)
\delta ^{ij}~
\left( \xi _{a,a}+{1\over 2}\hc _{aa}-{1\over 2}b_{aa}\right)\cr 
&-4\tilde \mu _s\Bigl( \xi _{i,j}+\xi _{j,i}+\hc _{ij}-b_{ij}\Bigr) \Bigr] .\cr
}\eqn\maamdetwo$$

\chapter{\bf Coupling to Gravity}

We choose Newtonian gauge (which is equivalent to the gauge
invariant formalism of Bardeen), so that the metric takes the form
$$ds^2=a^2(\eta )~\Bigl[ 
-d\eta ^2\bigl( 1+2\phi \bigr) +dx^idx^j
\left\{ \delta _{ij}~\bigl( 1-2\psi \bigr) 
+\hc ^{(V)}_{ij} +\hc ^{(T)}_{ij}
\right\} ~\Bigr] \eqn\ptaa$$
where $\hc ^{(V)}_{ij}$ and $\hc ^{(T)}_{ij}$
are pure {\it vector} and {\it tensor} parts of the 
spatial-spatial metric perturbation, respectively.
In dealing with cosmological perturbations it is convenient
to define any vector or tensor that can be expressed by 
taking derivatives of a scalar quantity as {\it scalar.}
Likewise, a tensor that can be expressed as a derivative
of a {\it vector} is regarded as {\it vector.}
With these definitions the linearized equations 
separate into independent {\it scalar,} {\it vector,}
and {\it tensor} blocks.
We assume a flat universe filled with cold dark matter and a
solid dark matter component.

We decompose the displacement field
$$\xib =\xib ^{(S)}+\xib ^{(V)},\eqn\ptaaab$$
and the internal metric of the solid dark matter component
$$b_{ij}=2b_{tr}^{(S)}~\delta _{ij}+
b_{ntr}^{(S)}\left( {k_ik_j-{1\over 3}\delta _{ij}k^2\over k^2}\right) 
+b_{ij}^{(V)}+b_{ij}^{(T)}.\eqn\ptaaa$$

\section{Scalar Perturbations}

We first consider the {\it scalar} perturbations. 
The linearized Einstein equations for the {\it scalar} sector are:
\relax{\eqnameb\boo  }
$$\eqalignno{
\delta {G_0}^0&={-2\over a^2}\cdot 
\left[ ~\nabla ^2\psi -3\H (\dot \psi +\H \phi )~\right] \cr
&=(8\pi G) \left[ ~ -\rho _c~\delta +{\Theta _0}^0~\right] ,&\boo a)\cr
\delta {{G^{(S)}}_i}^0&={-2\over a^2}\cdot 
\left[ ~\dot \psi +\H \phi ~\right] _{\vert i}\cr
&=(8\pi G) \left[ ~\rho _cv_{\vert i}+{{\Theta ^{(S)}}_i}^0~\right] , &\boo b)\cr
\delta {{G^{(S-tr)}}_i}^j&={2\over a^2}\cdot \left[ ~
\ddot \psi +2\H \dot \psi +\H \dot \phi +(2\dot \H +\H ^2)\phi +
{1\over 3}\nabla ^2(\phi -\psi )~\right] ~{\delta _i}^j\cr
&=(8\pi G)~{{\Theta ^{(S-tr)}}_i}^j,&\boo c)\cr
\delta {{G^{(S-ntr)}}_i}^j&={-1\over a^2}\cdot \left[ ~ \left( 
\nabla _i\nabla ^j-{1\over 3}{\delta _i}^j~\nabla ^2\right) 
(\phi -\psi )\right] ~ \cr
&=(8\pi G)~{{\Theta ^{(S-ntr)}}_i}^j&\boo d)\cr }$$
where $tr$ and $ntr$ denote the pure trace and traceless {\it scalar} parts of
the spatial-spatial tensors, respectively. The dots represent 
derivatives with respect to conformal time, $\H =(\dot a /a),$
$\rho _c=\Omega _{cdm}~(3/8\pi G)~\H ^2 a^{-2},$ and $v_i=v_{\vert i}$
(i.e., $v$ is the potential for
the {\it scalar} part of the velocity field).
The covariantly divergenceless tensor $\Theta _{\mu \nu }$
is the perturbation in the stress-energy of the solid dark matter component.

\def\V{{\cal V}}

Eqns.~\boo a) and \boo b) may be combined to obtain 
$$\nabla ^2 \psi ={3\over 2} \H ^2 \left[ ~
\Omega _{cdm}~\Bigl( ~\delta _{cdm}-3\H ~v_{cdm}~\Bigr)
+{1\over \rho _{crit}}\Bigl( -{\Theta _0}^0-3\H \V \Bigr) 
\right] \eqn\spb$$
where ${\Theta ^0}_i(${\it string-scalar}$)=\V _{\vert i}$
(i.e., \V ~is the potential used to represent the {\it scalar} component
of the solid dark matter component momentum density).
Similarly, eqn.~\boo d) may be recast as
$$\left( \nabla _i\nabla ^j-{1\over 3}{\delta _i}^j~\nabla ^2\right)
\bigl( \psi -\phi \bigr)
=3\H ^2~{1\over \rho _{crit}}~{{\Theta ^{(S-ntr)}}_i}^j.\eqn\spc$$
The equations of motion for the cold dark matter (CDM) component are
$$\eqalign{
&\dot \delta _{cdm}=-(\nabla \cdot {\bf v}_{cdm})+3\dot \psi ,\cr 
&\dot {\bf v}_{cdm}+\H ~{\bf v}_{cdm}=-\nabla \phi .\cr
}\eqn\spd$$
For the {\it scalar} mode of the solid dark matter component,
the equation of motion is
$$\eqalign{
&(1-\tilde \tau _s)
\Bigl\{ \ddot \xib ^{(S)}+(\lambda +1)\H ~ \dot \xib ^{(S)}
+\nabla \phi \Bigr\} 
= \left( 2\tilde K_s+{16\tilde \mu _s\over 3}\right) \Bigl\{ \nabla ^2\xib ^{(S)}-
3\nabla \left( \psi +b^{(S)}_{tr}\right) \Bigr\} .\cr } \eqn\spe$$
Finally, we have the equations
$$\eqalign{
{\Theta _0}^0&={-(\rho _s-\tau _s)\over a^{(3-\lambda )}}~\left[ 
-\xi _{i,i}-{1\over 2}\hc _{ii}+{1\over 2}b_{ii}\right] \cr 
&={-(\rho _s-\tau_s)\over a^{(3-\lambda )}}~
\left[ -\xi _{i,i}+3\psi +3b_{tr}^{(S)}\right],\cr 
{{\Theta ^{(S-ntr)}}_i}^j&= -4\tilde \mu _s{\rho _s\over a^{(3-\lambda )}}
\left[ \xi _{i,j}+\xi _{j,i}-
{2\over 3}\delta _{ij}\xi _{k,k}-b_{ij}^{(S-ntr)}\right] .\cr }
\eqn\spff$$

With the equations of motion for the 
cosmological perturbations including the solid dark matter component
derived, we now turn to initial conditions.
For each wavenumber ${\bf k}$ there exist four modes: two
growing modes and two decaying modes. This pair of perturbations
corresponds to the two distinct ways in which the solid dark matter generates
and alters the growth of perturbations.  If prior to the phase
transition that produced the solid dark matter, there were 
pre-existing curvature fluctuations, then the presence of the solid dark 
matter alters
the evolution of the fluctuations.  This is similar to the way 
in which a cosmological constant, neutrinos, or quintessence alters 
the evolution of the fluctuations.  In addition, the generation of 
the solid dark matter (e.g., string
formation) at a phase transition can produce new fluctuations.  
These entropy fluctuations correspond to variations in the dark matter 
density at the surface $T=T_{pt}$, where $T_{pt}$ is the phase 
transition temperature.  These
white noise entropy fluctuations are likely 
to be small on scales large compared to
the horizon scale at the phase transition.

We focus on the effect of the solid dark matter on the evolution of
pre-existing scalar, vector and tensor fluctuations.  Inflation
generates primarily scalar and tensor fluctuations; however, 
we include the vector term for
completeness.
On the surface $T=T_{pt}$ the solid dark matter component
inherits as its intrinsic 
spatial metric the metric on this surface
induced by the background spacetime metric.
Specifically, for small perturbations this gives
$$ b_{ab}=\left[ -2\psi +{2\over 3}{\delta _{rad}\over (1+w_{rad})}
\right] ~\delta _{ab},
\eqn\spfg$$
or, equivalently,
$$b_{tr}^{(S)}=-\psi +{1\over 3}{\delta _r\over (1+w_r)},
\qquad b_{ntr}^{(S)}=0.
\eqn\spfgga$$
Initially, we assume that $\xib =\dot \xib =0.$
The second term in \spfg ~arises from the shift in time of the 
surface of constant density relative to the surfaces
of constant cosmic time for Newtonian gauge. Since
the wavelengths of interest at this point lie 
far beyond the Hubble length, we have ignored 
perturbations in the velocity of the radiation
fluid. We assume that $(k\eta )\ll 1$ and that 
only the growing mode is relevant. 

The perfect fluid analogue of the above is 
as follows. For temperatures $T>T_{pt},$ the universe
is filled with a single perfect fluid, which at
$T=T_{pt}$ instantaneously splits into several
uncoupled perfect fluid components, labeled by 
$(A=1, \ldots , N).$ In this case the matching
condition is
${\delta /(1+w)}={\delta _A/(1+w_A)}$ 
for all A and all velocities may be neglected.
While the Lagrangian formalism developed in this
paper rather than the more familiar Eulerian formalism
could be used to describe this perfect fluid situation,
for non-Abelian strings, and similarly any solid with 
harmonic resistance to shear, the more general 
Lagrangian description is required.

We now consider the subsequent evolution given these
initial conditions. In the situations of interest 
the solid dark matter component is formed well
before radiation-matter equality and the 
solid dark matter component contributes negligibly to
the matter density of the universe compared
to other components until well into the 
matter dominated epoch. 

Assuming either complete matter or complete radiation domination
gives $\delta =-2\phi =-2\psi $ on superhorizon scales. Consequently,
$b_{tr}^{(S)}=-(3/2)\psi $ during radiation
domination on the scales of interest. During the matter-radiation
transition $\psi $ drops by a factor of 
$(9/10)$ while $b_{tr}^{(S)}$ does not change;
therefore, during matter domination
on superhorizon scales $b_{tr}^{(S)}=-(5/3)\psi .$

To follow the evolution of the perturbations
through the transition from matter
to solid dark matter domination, it is convenient to 
define the variable $S=(\nabla \cdot \xib ).$ 
The equation of motion becomes
$$\eqalign{
&(1-\tilde \tau _s)\ddot S+(1+\lambda )\H (1-\tilde \tau _s)\dot S
-\left( 2\tilde K_s +{16\tilde \mu _s\over 3}\right)
\nabla ^2S\cr
&\hskip 1in =-\left[
(1-\tilde \tau _s)\nabla ^2\phi
+3\left( 2\tilde K_s +{16\tilde \mu _s\over 3}\right) 
\nabla ^2\left( \psi +b_{tr}^{(S)}\right) \right] ,
\cr }\eqn\sggb$$
and the sources become 
$$\eqalign{
{\Theta _0}^0&= \rho _{crit}~\Omega _{sdm}~
(1-\tilde \tau _s)~\Bigl[ S-3\left( \psi +b_{tr}^{(S)}\right) \Bigr] ,\cr
{{\Theta ^{(S-tr)}}_i}^j&= {\delta _i}^j~(-2\rho _{crit}~\Omega _{sdm}~
\tilde K_s)~\Bigl[ S-3\left( \psi +b_{tr}^{(S)}\right) \Bigr] ,\cr 
{{\Theta ^{(S-ntr)}}_i}^j&= -8\tilde \mu _s~\rho _{crit}~\Omega _{sdm}~
\left( {k_ik^j-{1\over 3}k^2{\delta _i}^j\over k^2}\right) ~S.\cr
}\eqn\sggc$$
It follows that
$$-\nabla ^2(\psi -\phi )=k^2(\psi -\phi )=24\tilde \mu _s~\Omega _{sdm}~\H ^2S
\eqn\sgge$$
and 
$$\eqalign{
&\ddot \psi +2\H \dot \psi +\H \dot \phi +
(2\dot \H +\H ^2)\phi +{1\over 3}\nabla ^2(\phi -\psi )\cr
&\hskip 1in
=-3\H ^2 \Omega _{sdm}\tilde K_s~
\Bigl[ S-3(\psi +b_{tr}^{(S)})\Bigr] .\cr }\eqn\sggm$$
Initially, far outside the horizon, $S=\dot S=0.$ It follows that
$S=O(1)\cdot (k\eta )^2\cdot \psi $ on superhorizon scales.

The evolution of the gravitational potentials after the radiation epoch
may be computed by solving 
equations \sggb , \sgge , and \sggm ~combined with the initial
conditions $S=\dot S= 0,$ $\psi =\psi _{init},$ $b_{tr}^{(S)}=-(5/3)~
\psi _{init},$ and $\dot \psi =0.$ For a non-Abelian string network
stretched by the Hubble flow, the evolution of the scale factor
is given by $a(\eta )=\bar a ~[ \cosh \eta -1 ]$ 
where $\Omega _m=\sech ^2[\eta /2],$ just as for a hyperbolic
universe with CDM.

\mathchardef\etab="0911

As a practical matter, it is better to use synchronous 
gauge to evolve the perturbations because synchronous 
gauge is better behaved on superhorizon
scales. In Newtonian gauge the density and velocity 
perturbations on superhorizon scales are not small. 
This results from the warping of surfaces of constant
cosmic time required to make the spatial part
of the metric conformally flat. As a consequence using the 
constraint equations to determine the Newtonian potentials
and their time derivatives involves delicate cancellations
between large quantities, cancellations that become
increasingly delicate as one passes to earlier times. 
Synchronous gauge, on the other hand, is more robust.
For the adiabatic growing mode synchronous gauge (with the 
amplitude of the gauge modes set to zero) rapidly
approaches a co-moving, constant density gauge as one passes to 
superhorizon scales. With the convention\refmark{\cpma }
$$ ds^2=-d\eta ^2+a^2(\eta ) ~\Bigl[
\delta _{ij}+
h({\bf k},\eta ) \hat k_i\hat k_j+
6\etab ({\bf k},\eta )~\Bigl( \hat k_i\hat k_j-
{1\over 3} \delta _{ij} \Bigr) 
\Bigr] dx^i~dx^j,
\eqn\craa$$
at early times during the radiation epoch on 
superhorizon scales one has 
$\delta \sim O(k^2\eta ^2)$ and 
$\theta \sim O(k^4\eta ^3)$ for all 
components contributing to the stress-energy and
$h\sim O(k^2\eta ^2),$
and the only appreciable perturbation 
is $\etab \sim O(1).$
For the solid dark matter for initial conditions on
superhorizon scales it is an adequate approximation
to set $\xib =\dot \xib =0$ and 
$b_{ij}=6\etab ({\bf k},\eta )~\Bigl( \hat k_i\hat k_j-
{1\over 3}\Bigr) \delta _{ij}.$ 
The Newtonian potentials may be calculated from
the synchronous variables according to 
$$\eqalign{
\phi &= {1\over 2k^2}\Bigl[
(\ddot h+6\ddot \etab )+\H (\dot h+6\dot \etab )
\Bigr],\cr 
\psi &= \dot \etab -{\H\over 2k^2}
(\dot h+6\dot \etab ).\cr 
}\eqn\crab$$

\section{Vector Perturbations}

For completeness in this subsection we give the evolution equations
for the vector sector. Although for each wavenumber ${\bf k}$ the
solid dark matter component has two dynamical vector degrees of freedom,
for the usual inflationary models combined with a 
solid dark matter component these vector modes are not
excited. As before, on the initial surface at $T=T_{pt}$ the 
solid dark matter component
inherits as its intrinsic metric the metric
on this surface induced by the spacetime metric, but if
the $\hc ^{(V)}_{ij}=0,$ it follows that $b^{(V)}_{ij}=0.$
Similarly, on this surface on superhorizon scales
$\xib ^{(V)}=\dot \xib ^{(V)}=0.$ We also have the gauge condition
$\hc _{0i}^{(V)}=0.$ 

For the CDM the equation of motion for the {\it vector} modes is
$$ \dot {\bf v}_{cdm}^{(V)}+\H ~{\bf v}_{cdm}^{(V)}={1\over 2}\nabla 
\cdot \underline{\hc }^{(V)}.  \eqn\spee$$
Similarly, for the two {\it vector} modes of the solid dark matter component,
the equations of motion are
$$\eqalign{
&(1-\tilde \tau _s)\Bigl\{ \ddot \xib ^{(V)}+(1+\lambda )\H ~
\dot \xib ^{(V)}\Bigr\} 
= 2\tilde \mu \Bigl\{ \nabla ^2\xib ^{(V)}
+\nabla \cdot \underline{\hc }^{(V)}-\nabla \cdot \underline{b}^{(V)}
\Bigr\} .\cr }
\eqn\spef$$

The Einstein equations for the {\it vector} sector are
$$\eqalign{
{1\over a^2}\nabla _j\cdot \dot \hc ^{(V)}_{ji}&=
(8\pi G)~\Bigl[ \rho _{cdm}{v_{cdm}}_{~i}^{(V)}+\Theta ^{(V)}_{0i}\Bigr] ,\cr
{1\over a^2}\Bigl[ \ddot \hc ^{(V)}_{ij}+2\H \dot \hc ^{(V)}_{ij}\Bigr] &=
(8\pi G)~\Theta ^{(V)}_{ij}.\cr }
\eqn\speg$$

Finally, we have the equations
$$\eqalign{
{{\Theta ^{(V)}}_i}^0&={\rho _s\over a^{(3-\lambda )}}
(1-\tilde \tau _s)~\dot \xi _i^{(V)},\cr
{{\Theta ^{(V)}}_i}^j&={\rho _s\over a^{(3-\lambda )}}
(-4\tilde \mu _s )\left[ \xi ^{(V)}_{i,j}-\xi ^{(V)}_{j,i}
+\hc ^{(V)}_{ij}-b^{(V)}_{ij}\right] .\cr }
\eqn\spffgg$$

\section{Tensor Perturbations}

Primordial tensor perturbations, such as those generated during
inflation, are influenced by the solid dark matter component.

The linearized Einstein equation for 
the {\it tensor} sector is
$${1\over a^2}\Bigl[ \ddot \hc ^{(T)}_{ij} +2\H ~\dot \hc ^{(T)}_{ij}
-\nabla ^2 \ddot \hc ^{(T)}_{ij}\Bigr] =
(8\pi G)~\Theta ^{(T)}_{ij}. \eqn\gwa$$
The {\it tensor} stress-energy from the solid dark matter component 
is given by
$${\Theta ^{(T)}_i}^j=-4\tilde \mu _s~\rho _{crit}~\Omega _s ~
\Bigl[ {{\hc ^{(T)}}_i}^j - {{\hc ^{(T)}}_i}^j(\eta =0)\Bigr] ,
\eqn\gwab$$
which may be inserted into \gwa ~to obtain
$$\ddot \hc ^{(T)}_{ij} +2\H ~\dot \hc ^{(T)}_{ij}
-\nabla ^2 \ddot \hc ^{(T)}_{ij}+12\tilde \mu _s\Omega _{sdm}~
\H ^2\Bigl[ \hc ^{(T)}_{ij}-\hc ^{(T)}_{ij}(\eta =0)\Bigr] 
= 0.\eqn\gwac$$
Physically, the response of the solid dark matter component 
contributes a mass term to the gravity waves.

\def\gtorder{\mathrel{\raise.3ex\hbox{$>$}\mkern-14mu
             \lower0.6ex\hbox{$\sim$}}}
\def\ltorder{\mathrel{\raise.3ex\hbox{$<$}\mkern-14mu
             \lower0.6ex\hbox{$\sim$}}}

\chapter{Implications for the CMB Anisotropy}

We now explore the consequences of a solid dark
matter component for the predicted CMB anisotropy.
The {\it scalar} contribution is given by the Sachs-Wolfe
formula
$${\Delta T\over T}(\theta ,\phi )=
{1\over 3}\phi (r_{ls}, \theta ,\phi , \eta _{ls})
+\int _{\eta _{ls}}^{\eta _0} d\eta ~
\left( {\partial \psi \over \partial \eta }
+   {\partial \phi \over \partial \eta }
\right) \Bigg|_{r=\eta _0-\eta .}
\eqn\grrc$$
Expanding the CMB anisotropy into
spherical harmonics
$${\Delta T\over T}(\theta ,\phi )=\sum _{lm}
a_{lm}~Y_{lm}(\theta ,\phi ),\eqn\grrd$$
one obtains the following expression for the expected 
multipole moment variance:
$$c_l=\langle \vert a_{lm}\vert ^2\rangle 
=\int _0^\infty dk~P(k)~\left[ {1\over 3}
\Phi (\eta _{ls};k)j_l(kr_{ls})
+\int _{\eta_{ls}}^{\eta _0}
d\eta ~j_l(kr)~\biggl\{ \dot \Psi (\eta , k)+
\dot \Phi (\eta , k)\biggr\} 
\right] ^2.\eqn\grre$$
Here $P(k)$ is the primordial power spectrum, 
which we set to $P(k)\prop 1/k,$ 
indicating a featureless, scale-invariant,
Harrison-Zeldovich primordial spectrum. Here
the functions $\Psi (\eta , k)$ and $\Phi (\eta , k)$
indicate the time dependence of the 
growing modes of wavenumber $k$ of
$\psi $ and $\phi ,$ respectively, and are normalized
to unity as $\eta \to 0.$

On small angular scales $(\ell \gtorder 30),$
the solid dark matter component does not play
a significant role in determining the CMB
anisotropy because on these scales the 
anisotropy is almost exclusively determined
by what happens at the surface of last scatter
when $\Omega _{solid}$ is negligible, and the
contribution of the integrated Sachs-Wolfe term
on these scales is negligible. On larger
angular scales, however, the contribution
through the integrated Sachs-Wolfe term of
the decay of the gravitational potential
contributes significantly to the low-$\ell $
moments, and since the details of how the
potential decays depend on the dynamics
of the smooth component, one expects the
dynamics of the solid dark matter component
to play an important role in determining these CMB moments.

To illustrate the effect of a solid dark matter component, 
we compare the evolution of the gravitational potentials
and the large-angle CMB moments for the following nine
cosmological models, some with a solid dark matter component
and others included for purposes of comparison:

{\narrower

\underbar{0. A flat $\Omega _m=1$ universe.} In this critical
universe filled with cold dark matter the gravitational potentials
$\phi $ and $\psi $ remain constant at late times, so there
is no integrated Sachs-Wolfe contribution.

\underbar{1. A hyperbolic $\Omega _m=0.3$ universe.} The scale
factor $a(\eta )$ for this hyperbolic universe with no dark matter other
than a subcritical cold dark matter component evolves exactly
as the scale factor for the string dominated universe. However,
the negative spatial curvature and differing decay of the
gravitational potential at late times leads to a different
shape for the low-$\ell $ CMB moments.

\underbar{2. A flat $\Omega _m=0.3$ string dominated universe.} In 
this flat universe with a subcritical density of cold dark matter
and the remainder in a network of frustrated non-Abelian strings,
the physical density in the solid string network component scales
as $\rho _s\prop a^{-2},$ becoming the dominant form of matter
at late times. The behavior of the solid string component depends
on the strength of the resistance to pure shear $\tilde \mu _s.$
We consider the following three subcases:\hfill \break
\hbox to .75in {}\underbar{2a. $c_S=0,$ $c_V=1/2.$} 
($\tilde \mu _s=1/24$).\hfill \break
\hbox to .75in {}\underbar{2b. $c_S=1/\sqrt{3},$ $c_V=1/\sqrt{2}.$}
($\tilde \mu _s=1/12$).\hfill \break
\hbox to .75in {}\underbar{2c. $c_S=1,$ $c_V=1.$}
($\tilde \mu _s=1/6$).

\underbar{3. A flat $\Omega _m=0.3$ domain wall dominated universe.} 
In this flat cosmological model a network of frustrated domain walls
formed in a late-time phase transition gives a density that scales
as $\rho _s=1/a.$ As for the string-dominated universe, we consider
three subcases:\hfill \break 
\hbox to .75in {}\underbar{3a. $c_S=0,$ $c_V=1/\sqrt{2}.$}
($\tilde \mu _s=1/24$).\hfill \break
\hbox to .75in {}\underbar{3b. $c_S=1/\sqrt{3},$ $c_V=\sqrt{3}/2.$}
($\tilde \mu _s=3/48$).\hfill \break
\hbox to .75in {}\underbar{3c. $c_S=\sqrt{2/3},$ $c_V=1.$}
($\tilde \mu _s=1/12$).

\underbar{4. A flat $\Omega _m=0.3$ $\Lambda $-dominated universe.}
This flat universe with a cosmological constant may be interpreted
as a degenerate case of a solid dark matter component in
which $\tau _s\to 1.$

}

\hskip 30pt

Figures 1(a) and 1(b) indicate the evolution of the gravitational potentials
$\phi $ and $\psi ,$ respectively, in the limit $k\to 0$ (i.e., on 
superhorizon scales) for the various models. The potentials have 
been normalized to unity at early times and the horizontal axis
indicates conformal time, normalized so that $\eta =1$ today.
Models 1, 2a-c, 3a-c, 4 are indicated, with no offset for model 1 and 
offsets increasing by $.1$ for each successive model, introduced to separate the
curves in the plot. 
Although models (1) and models (2a)-(2c) have the
same evolution of the scale factor, the evolution
of the potential is different at later times.
The fact that $\phi $ and $\psi $ evolve differently
is due to the presence of large anisotropic stresses.
In the domain wall dominated models the decay
of the Newtonian potential $\psi $ is much greater
than in the hyperbolic or $\Lambda $ models leading
to a significantly larger integrated Sachs-Wolfe
contribution to the CMB moments.

Figure 2 shows the CMB moments for these models from $\ell =2$
through $\ell =30$ normalized so that $c_{30}$ agrees for all
the models. The vertical axis is $c_l\cdot \ell (\ell +1)$
with offsets increasing by of $.2$
The shapes of the CMB moments were computed with a method
that does not take into account the effects
producing the beginning of the rise toward the Doppler
peak toward increasing $\ell .$ In other words,
$\ell (\ell +1) c_\ell $ would be constant for a flat
CDM universe when in fact there is a
40 \% rise in this quantity by $\ell =30.$ Therefore, only
the relative differences in shape are signficant. A more
detailed study of these models
using a Boltzman code will be presented in
a forthcoming paper.

\chapter{\bf Discussion}

In this paper we have developed a continuum formalism 
for describing the dynamics of a `solid dark matter' (SDM)
component and shown how cosmological perturbations
evolve with such a component included. The advantages
of positing an SDM component are: (1). It is possible
to reconcile a spatially flat cosmology with the 
numerous measurements of $\Omega $ indicating a low
value because most methods of measuring $\Omega $
are sensitive only to matter that is clustered
(e.g., on scales comparable to the size of
galaxy clusters or smaller) and the SDM remains
unclustered, except on the very largest scales
comparable in size to our present
horizon. (2). SDM can provide the negative pressure
suggested by the recent SNIa observations at
high redshift, thus explaining the lower than
expected apparent luminosities of the
distant supernovae.  Supernova observations can potentially constrain
the equation of state, thus distinguishing a SDM-dominated universe
from a cosmological constant dominated universe. (3). With SDM it is not necessary to posit 
a new, surprisingly small energy scale. SDM
from string or domain wall networks results
from new physics at higher energy scales. 
SDM thus avoids the fine tuning difficulties of 
a straightout $\Lambda $ term or of the 
`quintessence' models with an extremely slowly 
evolving scalar field that gives the same 
qualitative effect as $\Lambda .$ 

Introducing an equation of state with negative pressure
is a delicate matter. If one wishes to consider perturbations
to homogeneity and isotropy, considerations of general covariance
and causality prohibit one from introducing a smooth
background component that does not cluster in an 
{\it ad hoc} way.
Since locally it is impossible to
determine what the `unperturbed'
background solution in the absence of perturbations would
have been, a `smooth component' must be introduced in a
manner that specifies how perturbations evolve,
and to do this more than merely specifying 
$w=(p/\rho )$ is required.
For an equation of state with negative pressure, 
to posit a perfect fluid is not allowed, because the 
sound speed on short wavelengths would be imaginary,
indicating instabilities whose 
growth rate diverges as the wavelength
approaches zero. In SDM a sufficiently
large shear modulus removes this
instability. In the slowly rolling field models
the instability is lacking for an entirely different
reason: there is an inertia associated with
changing the stress-energy. 

Physically, how the instability
is avoided in SDM and in quintessence is manifested
by the following qualitative differences: (1). SDM,
unlike quintessence and most types of dark
matter, generates anisotropic stresses. 
(2). SDM has vector modes with nonvanishing
sound speed. (3). The resistance to pure
shear in SDM gives the graviton a mass, 
changing the gravity wave contribution to
the CMB on large angular scales.  

We finally offer the following more technical
remarks comparing SDM to other possible sources 
of negative pressure discussed in the literature: 

{\narrower

\vskip 14pt 
\noindent (1). A pure cosmological constant may be regarded
as a degenerate case of the action \mact ~with
$\rho \bigl( g_{(1)},  g_{(2)},  g_{(3)} \bigr) $
set to $(\Lambda /8\pi )\sqrt {g_{ab}}.$
This degenerate case greatly enlarges the reparameterization 
invariance of \mact . Because $p=-\rho $ exactly, 
the smooth dark matter stress-energy 
no longer singles out a special time direction,
and consequently for this special case
general reparameterizations that mix $\y $ and $t$ are allowed.

\vskip 14pt 
\noindent (2). If $\mu =0,$ eqn.~\mact ~becomes a Lagrangian
description of a perfect fluid. In this special situation
the Lagrangian description is much more
cumbersome than the Eulerian description,
especially with general relativity taken into
account. When $\mu \ne 0,$ however, an 
Eulerian description is no longer possible.
If $\mu =0$ and $p=w\rho $ where $w<0$ and $w\ne -1,$
the sound speed becomes imaginary, indicating
an instability, most severe on the smallest
scales. Without resistance against pure shear,
the solid dark matter component would be similarly
unstable. However, for the non-Abelian strings
and domain wall networks
$\mu $ is sufficiently large to stabilize the medium,
as evidenced by all sound speeds being real. 

\vskip 14pt 
\noindent (3.) Quintessence differs in that the 
medium has an internal scalar degree of freedom.
One could in fact write down a more general action
for a low-energy description that combines 
quintessence with the continuous medium:
$$\eqalign{S&=-\int dt \int d^3y~\sqrt{h}~
\rho \bigl( g_{(1)},  g_{(2)},  g_{(3)}, \phi \bigr)
\sqrt{-G_{\mu \nu }
{\partial X^\mu \over \partial t}
{\partial X^\nu \over \partial t}}\cr 
&~~~+\int d^4x \sqrt{\vert G_{\mu \nu }\vert }~
{1\over 2}\left[ (\D _{\hat t}\phi )^2
-c_s^2\Bigl(\phi , g_{(1)},  g_{(2)},  g_{(3)}\Bigr) (\D _{\y }\phi )^2
\right] -V[\phi ].\cr }
\eqn\mactr$$
Because the stress-energy of the medium breaks Lorentz invariance
down to the rotation group, $c_s$ need not equal the speed of light.

4. Alexander Vilenkin\refmark{\avthree }
has pointed out that for the special
case $w=-1/3$ (i.e., frustrated strings), due to a
cancellation in Newtonian gauge the strings do not 
experience a gravitational force from nonrelativistic
matter. In eqn.~\moabb ~this can be seen 
as a cancellation between the gradients of $h_{00}$ and $h_{ii}$
that occurs only for this special case under the assumption that the
source of these potentials has no anisotropic stresses.

}

\noindent {\bf Acknowledgements.}
We would like to thank Curt Callan, Brandon Carter, Patrick McGraw, Paul
Steinhardt, Neil Turok, and Alexander Vilenkin for useful discussions
and especially Richard Battye for helping us compare our calculations
to those using CMBFAST.
MB was supported in part by National Science Foundation grant
PHY9722101 and by the UK Particle Physics and Astronomy Research 
Council. DNS acknowledges the MAP project for support.

\refout 

\centerline{\bf Figure Captions}

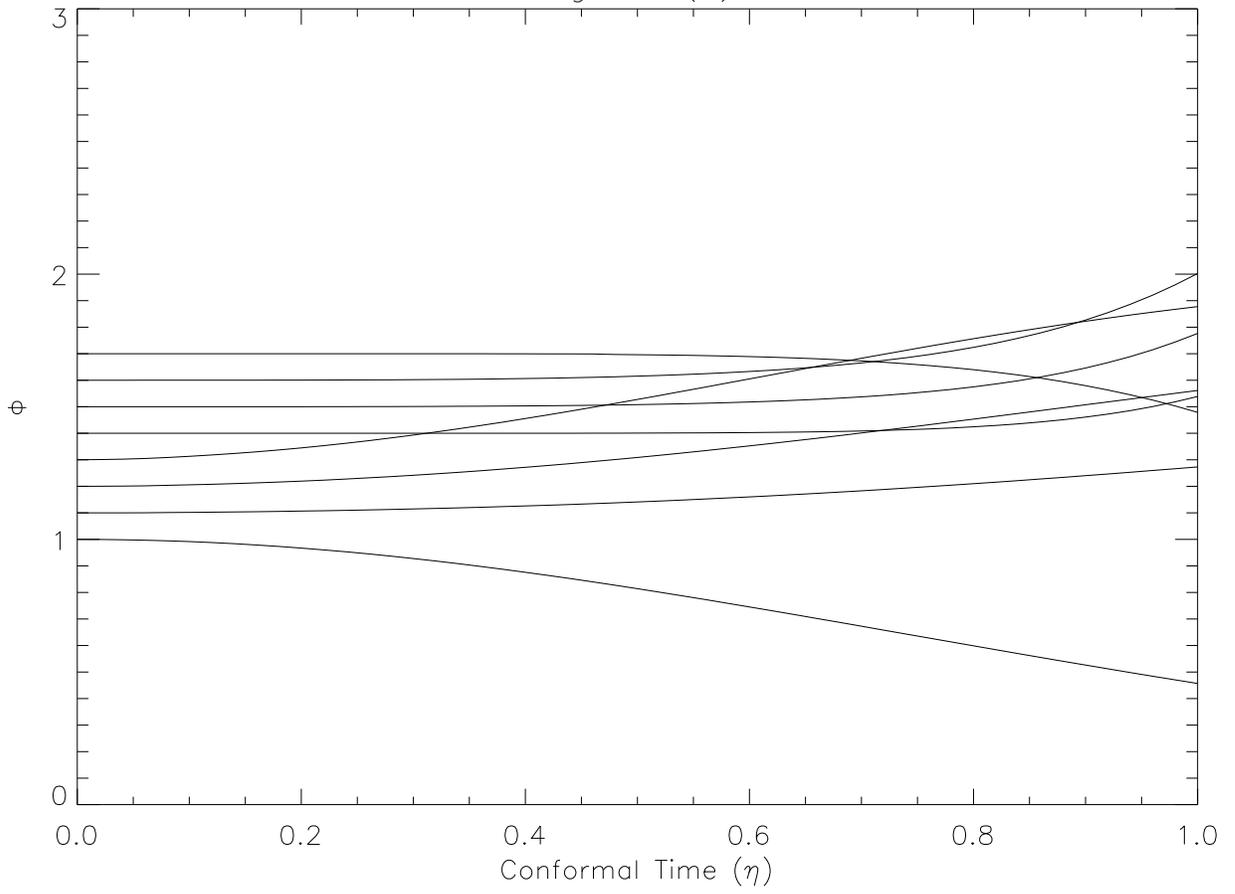
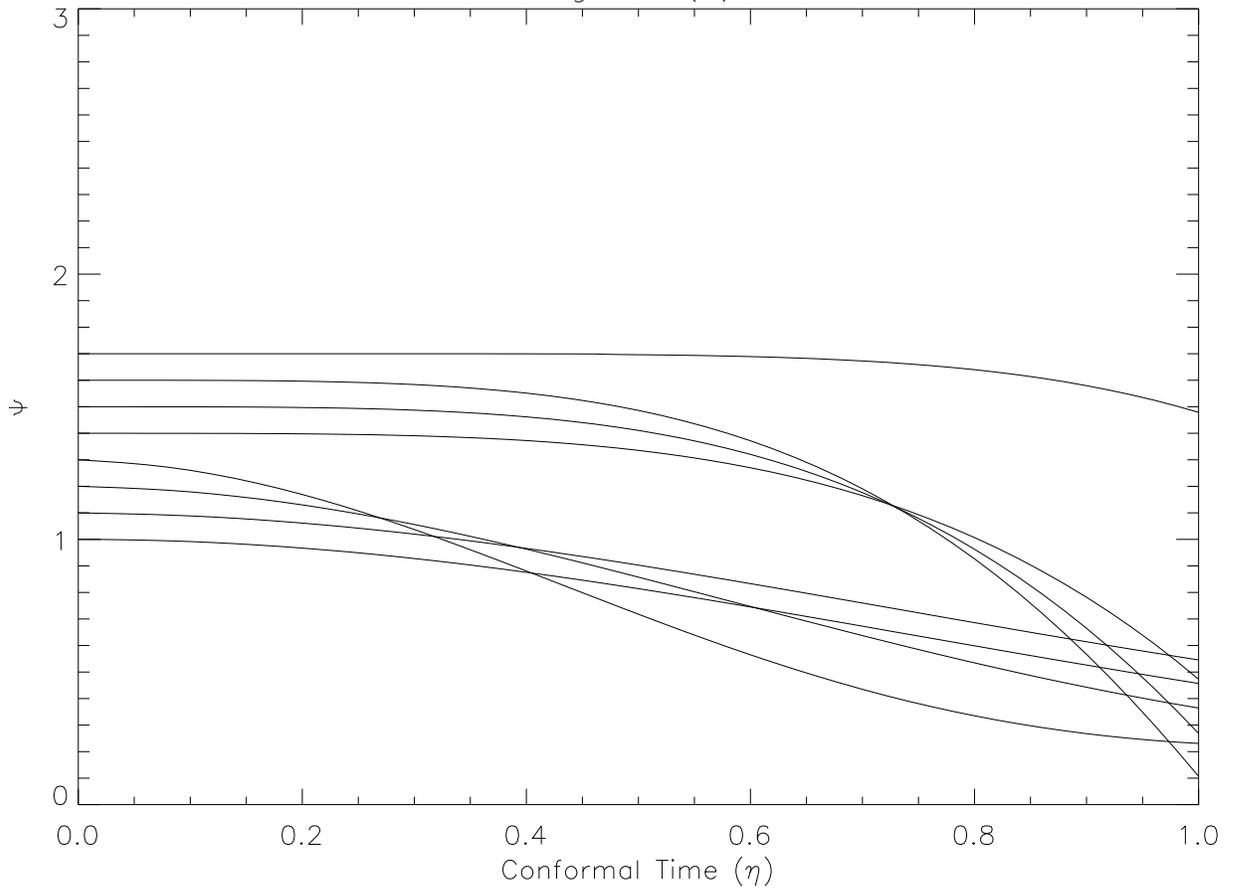
\item {Figure~1.} Panels (a) and (b) show the evolution of the potentials
$\phi $ and $\psi ,$ respectively, for vanishing wavenumber as a function
of conformal time for models 1, 2(a-c), 3(a-c), and 4. All the 
potentials are normalized to unity at vanishing conformal time. 
The curves for each successive model have been displaced upward 
by $0.1.$

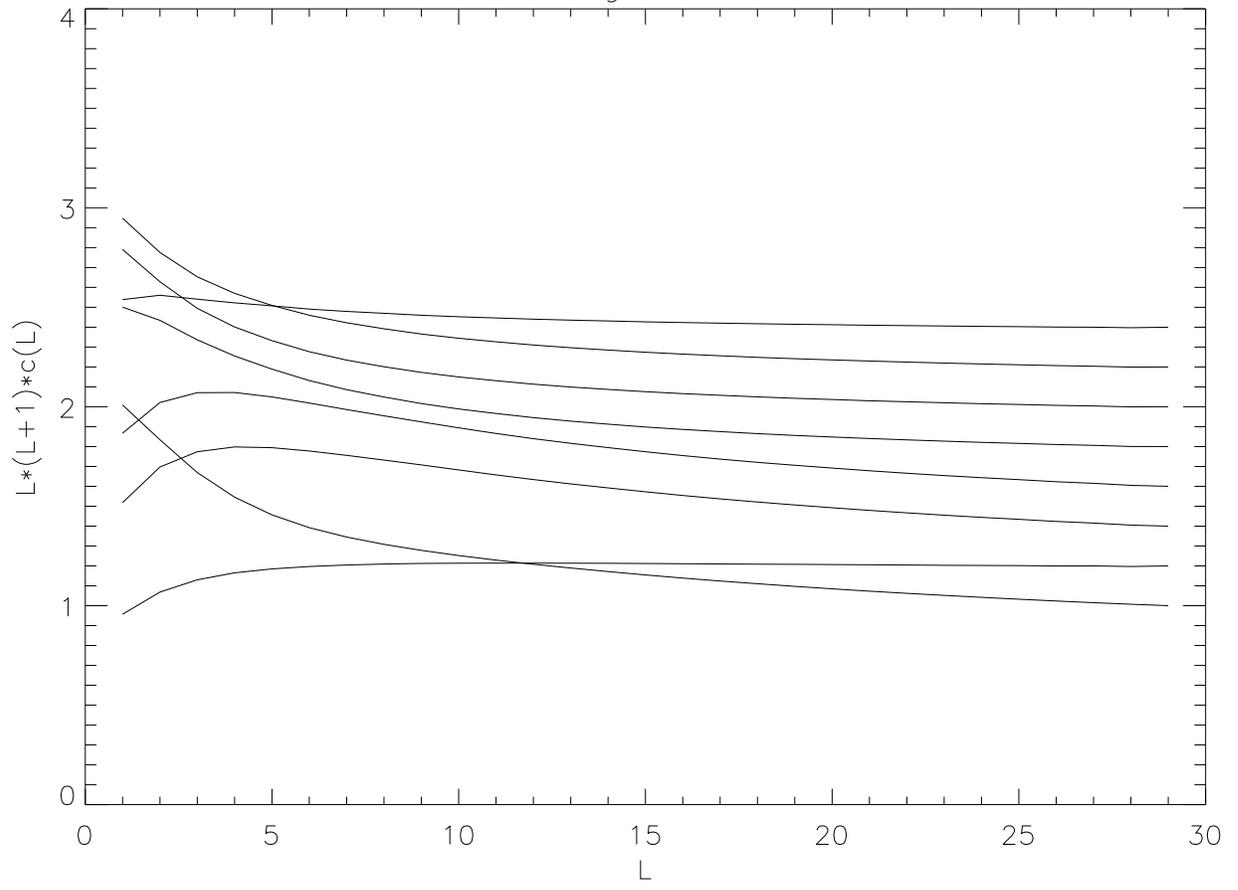
\item {Figure~2.} {\it CMB Multipole Moments.} The CMB multipole
moments are plotted for models  1, 2(a-c), 3(a-c), and 4 with 
successive curves displaced upward by $0.2.$ $\ell (\ell +1) c_\ell$
is plotted and the curves are normalized to unity at $\ell =30.$

\end